\documentstyle[prb,aps,epsf]{revtex}
\begin{document}
\draft

\title{Transfer of Spectral Weight in Spectroscopies of Correlated
Electron Systems}

\author{M. J. Rozenberg$^\dagger$, G. Kotliar$^\ddagger$ and
H. Kajueter$^\ddagger$}

\address{
$\dagger$ LPS Ecole Normale Sup\'{e}rieure, 24 rue Lhomond,
75231 Paris Cedex 05, France \\
$\ddagger$ Serin Physics Laboratory, Rutgers University,
Piscataway, NJ 08855-0849, USA}
\date{\today}

\maketitle

\begin{abstract}

We study the  transfer of spectral weight in the
photoemission and
optical spectra of strongly  correlated
electron systems.
Within the LISA,
that becomes exact in the limit of large
lattice coordination,  we consider and compare
two models of correlated electrons,  the Hubbard model and the
periodic Anderson model.
The results are discussed in regard of recent
experiments.
In the Hubbard model,
we predict an anomalous enhancement optical  spectral weight
as a function of temperature in the correlated metallic state
which is in qualitative agreement with
optical  measurements
in $V_2O_3$. We argue that
anomalies observed in the spectroscopy of
the metal are connected to
the proximity to a crossover region in the phase diagram
of the model.
In the insulating phase, we obtain an excellent agreement with the
experimental data and present a detailed
discussion on the role of magnetic frustration by studying the
$k-$resolved single particle spectra.
The results for the periodic Anderson model
are discussed in connection to recent experimental data
of the Kondo insulators
$Ce_3Bi_4Pt_3$ and $FeSi$. The model can successfully explain the different
energy scales that are associated to the thermal filling of the
optical gap, which we also relate to corresponding changes
in the density of states.
The temperature dependence of the optical sum rule is obtained
and its relevance for the interpretation of the experimental data
discussed. Finally, we argue that the large scattering rate measured
in Kondo insulators cannot be described by the
periodic Anderson model.

\end{abstract}
\narrowtext
\twocolumn

\section{\bf Introduction.} The interest in the
distribution of spectral weight in the
optical conductivity of correlated electron systems has been revived by
the improvement of the quality of the experimental data
in various systems \cite{thomas,bucher,schlesinger}.

The traditional methods used in the strong correlation problem,
exact diagonalization of small clusters \cite{sawatsky},
slave boson approaches\cite{millis}, and perturbative
calculations, have not been very successful in
describing the interesting
transfer of optical weight which
takes place as a function of temperature
in the strong correlation regime.

Recently, much progress has been achieved by mapping lattice models
into impurity models embedded in an effective medium.
This technique,
the Local Impurity Selfconsistent Approximation (LISA) \cite{review},
is a dynamical mean field theory that becomes exact in the limit
of large number of spatial dimensions \cite{vollmetz}.
For instance, the Hubbard and the Anderson lattice model can be mapped
onto the Anderson impurity model subject to different selfconsistency
conditions for the conduction electron bath \cite{gk,gkq}.
These resulting selfconsistent
impurity problems can be analyzed by a
variety of numerical techniques
\cite{mj,rzk,aw,zrk,pcj,agwk,rkz,ck,srkr,proj}.

In this paper we apply this approach to the study of the optical conductivity
in regard of the recent experiments in $V_2O_3$, $Ce_3Bi_4Pt_3$, and $FeSi$.
We take the view that the low energy optical properties of $V_2O_3$ can
be modeled by a one band Hubbard model, while
$Ce_3Bi_4Pt_3$ and $FeSi$ are described by a periodic Anderson model
\cite{aepplifisk}. The modeling
of the experimental systems
requires a large value of the Coulomb repulsion $U$.

Our main goal in this work is to
demonstrate that simplified models of strongly interacting systems treated
with the LISA, can account for the main qualitative
features that are observed experimentally in strongly correlated electron
compounds.
The paper is organized as follows: in section \ref{met} we summarize the
mean field equations for the model hamiltonians and the expressions for the
calculation of the optical conductivity and the optical sum rule.
In section \ref{phys} we present and intuitive pedagogical
discussion  the physical content of the solution
of the model hamiltonians in the large dimensional limit.
Section \ref{res} is dedicated to a thorough discussion of the optical
conductivity results.
We discuss the effects of magnetic frustration
on the spectral functions of
the Hubbard model.
And the effects of temperature and disorder on the optical
spectra of the Anderson lattice.
The theoretical calculations are carried out using exact
diagonalization (ED) and iterated perturbation theory (IPT) techniques,
and compared with texperimental results on various systems.
We stress that the use of IPT allow us to access physically
interesting regimes which are outside the scope of the Quantum
Monte Carlo method, and that the exact diagonalization technique is
used to confirm that the results presented are genuine features of
the Hubbard model in infinite dimensions and not artifacts of the IPT.
The conclusions are presented in the last section. Our
ED approach to the solution of
correlated models in  large dimensions
is based on the use  continuous fractions. The Appendix
describes  an  algorithm
to convert the sum of two given
continued fractions into a new continued fraction which we use
to extend the ED method to the models we treat in this paper.
Part of the theoretical results in section \ref{res} were announced
in a recent letter \cite{opt} . The optical conductivity
of the Anderson model and the Hubbard model were considered previously
by Jarrell {\it et al.} using the Qauntum Monte Carlo and Maximum
Entropy methods. \cite{pcj,mjam,cond}

\section{\bf Methodology.}
\label{met}
\subsection{Mean field equations.}
As model hamiltonians
we consider
the Hubbard and the periodic Anderson model (PAM):
\begin{equation}
H_H = -\sum_{<i,j>} (t_{ij}+\mu) c_{i\sigma}^{\dagger} c_{j\sigma}
 +  \sum_i U (n_{i \uparrow}-\frac{1}{2}) (n_{i\downarrow}-\frac{1}{2}),
\label{HubHam}
\end{equation}
\begin{eqnarray}
H_{PA}&=&\sum_{k} (\epsilon_{k} -\mu) c_{k\sigma}^{\dagger} c_{k\sigma}
+ \sum_i  (
{\epsilon}^o_d -\mu) d_{i\sigma}^{\dagger} d_{i\sigma}\nonumber \\
&+&\sum_i  V  d^{\dagger}_{i\sigma}
c_{i\sigma} + h.c.
+
 \sum_i U (n_{di \uparrow}-\frac{1}{2})
(n_{di\downarrow}-\frac{1}{2})
\label{AndHam}
\end{eqnarray}
where summation over repeated spin indices is assumed.
$\mu$ is the chemical potential, and $t_{ij}$ is the hopping amplitude
between the conduction electron sites, which in the PAM results in
the band $\epsilon_k$. The $d^\dagger$ and $d$ operators create and
destroy electrons on localized orbital with energy $\epsilon_d^o$.
$V$ is the hybridization
amplitude between $c$ and $d-$sites, which also appear in the
literature as $d$ and $f-$sites respectively.

The derivation has been given in detail elsewhere \cite{gk,gkq}.
So we only present
the final expressions.
The resulting local effective action reads,
\begin{eqnarray}
{S}_{local}&=&
- \int^{\beta}_{0} d\tau \int^{\beta}_{0} d\tau'
\psi^{\dagger}_{\sigma}(\tau) {\cal G}_{0}^{-1}(\tau-\tau')
\psi_{\sigma}(\tau')\nonumber\\
& +&\  U \ \int^{\beta}_{0}d\tau (n_{\uparrow}(\tau)-{1\over2})
(n_{\downarrow}(\tau)-{1\over2})
\label{localaction}
\end{eqnarray}
where $\psi^\dagger_\sigma,\ \psi_\sigma$ correspond to a particular site, and
denote $c^\dagger_\sigma,\ c_\sigma$
in the Hubbard model, and
$\{c^\dagger_\sigma,d_{\sigma}^{\dagger}\},\ \{c_\sigma,d_{\sigma}\}$
in the PAM case. $n_\sigma$ corresponds to $n_{c\sigma}$ and $n_{d\sigma}$
respectively. Also note that Eq.\ref{localaction} defines the associated
impurity problem, with $\psi^\dagger_\sigma,\ \psi_\sigma$ being the operators
at the impurity site while the information on the hybridization with the
environment is implicitly contained in ${\cal G}_{0}^{-1}$.
Requiring that $G_{local}(\omega)=\Sigma_k G(k,\omega)$,
we obtain as
selfconsistency condition
\begin{equation}
 {\cal G}_{0}^{-1}(\omega)= \omega + \mu - t^2 \tilde G(\omega)
\label{s0}
\end{equation}
for the Hubbard model, and
\begin{equation}
 [{\cal G}_{0}^{-1}]_{cc}(\omega)= \omega + \mu - t^2 [\tilde G]_{cc}(\omega)
\end{equation}
with ${\cal G}_0$ explicitly given by
\begin{equation}
{\cal G}_{0}^{-1}(i\omega)= \left(
\begin{array}{cc}
i\omega-{t^2}[\tilde G]_{cc}(i\omega) & V\\
V & i\omega
\end{array}
\right)
\label{s1}
\end{equation}
for the PAM.
In both cases, $\tilde G$ is the ``cavity'' Green function
which has the information of the response of the lattice.

We will consider the symmetric case  with
$\mu = 0$ and $ \epsilon^o_d = 0$.
Moreover, we assume a semi-circular bare
density of states for the conduction electrons, $\rho^o
(\epsilon) = \sum_{k}\delta(\epsilon-\epsilon_{k})/ N_{sites} =
(2 / {\pi D}) \sqrt{1 - ( \epsilon /D)^2}$, with the half-bandwidth $D=2t$.
This density of states can be realized in a Bethe lattice
and also on a fully connected fully frustrated version of the model
\cite{zrk,agwk}.
In this case the ``cavity'' Green function
simply becomes $\tilde G = G$.
In the following we set the half-bandwidth $D=1$.
We use an exact diagonalization algorithm (ED) \cite{ck,srkr}
and an extension of the  second order iterative perturbation theory
(IPT)
to solve the associated
impurity problem \cite{zrk}.
We have checked that IPT and the ED method are in good agreement
for all values of the model parameters.
This results from the property of IPT to capture the atomic limit exactly
in the symmetric case\cite{zrk}.
We use extensively the IPT on the real axis
to scan through parameter space. A detailed comparison will be presented
elsewhere.

\subsection{Optical conductivity.}

The optical conductivity of a given system is defined by
\begin{equation}
\sigma(\omega) = {1\over {\cal V}\,\omega} \mbox{Im}\,
\int_0^\infty \langle[{\cal J}(t),{\cal J}(0)] \rangle e^{i\omega t} dt
\end{equation}
where ${\cal V}$ is the volume, ${\cal J}$ is the
current operator and $\langle$ $\rangle$ indicates an
average over a finite temperature
ensamble or over the ground state at zero temperature.
In general  $\sigma(\omega)$ obeys a version of the
f-sum rule \cite{mald,kohn},
\begin{equation}
\int_0^\infty \sigma(\omega) d\omega = {\pi \over {\cal V}}{\rm Im}
\langle [ P, {\cal J}] \rangle
\end{equation}
where $P$ is a polarization operator obeying
$\frac{\partial P}{\partial t} = {\cal J}$.

In a model which includes {\it all electrons and all bands}
the current operator
${\cal J}$ is given by
\begin{equation}
{\cal J}=\frac{e}{m} \sum_i  p_i \delta(r-r_i)
\end{equation}
where $p_i$ is the momentum and $r_i$ the position of the
${\rm i}^{th}$-electron, and
$e$ and $m$ denote its charge and bare mass.
$P$ is given by
\begin{equation}
P=e \sum_i r_i \delta(r-r_i).
\end{equation}
Thus, ${1\over {\cal V}} \langle [ P,{\cal J}] \rangle=\frac{i ne^2}{m}$
where $n$ is the density of
electrons, and the sum rule becomes
\begin{equation}
\int_0^\infty \sigma(\omega) d\omega = \pi \frac{n e^2}{m}.
\end{equation}
This result is clearly temperature independent and does not depend either
on the strength of the interactions.

When dealing with strongly correlated electron systems, in a frequency range
where few of the bands are believed to be important, it is
customary to work with an effective model with one or two bands, such as
the Hubbard or the periodic Anderson model.
The current operator is thus projected onto the low energy sector
and is expressed in terms of creation and destruction operators of the
relevant bands (i.e., ${\cal J} = i\frac{t}{\cal V}\sum_i
(c^\dagger_{i+\delta}c_i -c_i^\dagger c_{i+\delta})$
for the Hubbard and Anderson model).
It is in this case that the expectation value
$\langle [ P, {\cal J} ] \rangle $
is no longer $ \sim \frac{ne^2}{m}$, but,
proportional to
the expectation
value of the kinetic energy
$\langle K \rangle$
of the conduction electrons\cite{mald,rice}. In general $\langle K \rangle$
depends on
the temperature and strength of interactions, therefore, for these
few bands models, the optical weight sum rule will depend on them as well.
If the projection onto a few band model is valid, this result
also implicitly indicates that a portion of the
optical spectral weight (the weight
not exhausted by $\langle K \rangle$) is transferred to much higher energies,
that is, to the bands that were excluded by the projection to low energies.

In this paper we do not address the question of the validity of the
low energy projection onto a few band model. Instead we focus on the
consequences of this assumption on the redistribution of the optical
weight within a mean field theory that is exact in the limit of
large dimensions.
Our main conclusion is that there is a considerable
temperature dependence of the
integrated spectral weight appearing in  the sum rule.

In infinite dimensions, $\sigma(\omega)$ can be expressed in terms of the
one particle spectrum of the current carrying electrons \cite{khurana,cond}:
\begin{eqnarray}
\sigma(\omega)&=&
\frac{1}{\omega} \frac{2 e^2 t^2 a^2}{\nu \hbar^2}
\int_{-\infty}^{\infty}{d\epsilon}\ {\rho^o(\epsilon)}
\int_{-\infty}^{\infty}\frac{d\omega'}{2\pi}\nonumber \\
& & A_\epsilon(\omega') A_\epsilon(\omega'+\omega)
(n_f(\omega')-n_f(\omega'+\omega))
\label{sigma}
\end{eqnarray}
with
$A_\epsilon(\omega) = -2 {\rm Im}[G_k(\omega)]$
being the spectral representation of the Green function of the
 lattice conduction electrons,
 $a$ the lattice constant, and $\nu$ the
volume of the unit cell.

As we anticipated,
the kinetic energy is related to the conductivity by the sum rule
\begin{equation}
\int_0^\infty\sigma(\omega) d\omega = -\frac{\pi e^2 a^2}{2d\hbar^2 \nu}
\langle K \rangle =
\frac{\omega_P^2}{4\pi}
\label{sum}
\end{equation}

An important result, which will be demonstrated later on,
is the notable dependence of the plasma frequency $\omega_P$ with temperature.
This feature will be seen to emerge as consequence that
correlation effects generate small energy scales (e.g. the ``Kondo
temperature'' of the associated impurity). It is
the competition between the small
scales and the temperature that gives rise to an  unusual temperature
dependence
to the integrated optical spectral weight.

At $T=0$, the optical conductivity of a metallic correlated electron system
can be parametrized by \cite{kohn}
\begin{equation}
\sigma(\omega)=
\frac{{\omega^*_P}^2}{4\pi}\delta(\omega) + \sigma_{reg}(\omega)
\end{equation}
where the coefficient in front of the $\delta$-function is the Drude weight and
$\omega^*_P$ is the renormalized plasma frequency. In the presence of disorder
$\delta(\omega)$ is replaced by a lorentzian of width $\Gamma$.

Evaluating equation (\ref{sigma}) at $T=0$ one finds in mean field theory that,
\begin{equation}
\frac{{\omega^*_P}^2}{4\pi}=
\frac{2\pi e^2}{\hbar^2 \nu} Z \sum_k \big(
\frac{\partial \epsilon_k}{\partial k_x}\big)^2 \delta(\epsilon_k)
\end{equation}
where $Z$ is the quasiparticle weight.
For the Hubbard model in infinite dimensions the expression above further
simplifies, and it only depends on the density of states
\begin{equation}
\frac{{\omega^*_P}^2}{4\pi} = \frac{4\pi t^2 e^2 a^2}{\hbar^2\nu}Z\rho^o(0).
\end{equation}

\section{\bf Physical Content of the Mean Field Theory.}
\label{phys}
\subsection{Hubbard Model.}
\begin{figure}
\centerline{\epsfxsize=2.9truein
\epsffile{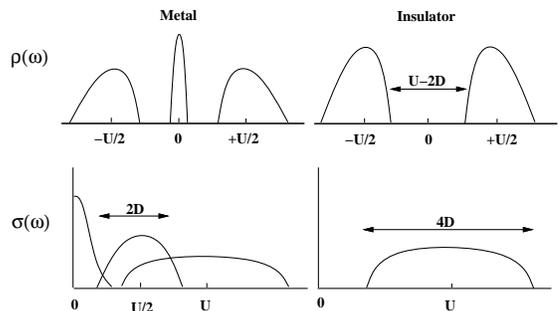}}
\caption{Schematic DOS for the Hubbard model (1/2 filling) and their
corresponding optical spectra for the metallic and insulator solutions.
The width of the incoherent peaks in the $DOS$ is $\approx 2D$ and the
one of central peak in the metal is $\approx ZD \equiv \epsilon_F^*$.
}
\label{fig1}
\end{figure}

The solution of the mean field equations shows that at low temperatures the
model has a metal insulator transition (Mott-Hubbard transition) at an
intermediate value of the interaction $U_c \approx 3D$ \cite{pcj,agwk,rkz}.
The metallic side is characterized by a density of states with a
three peak structure: a central
feature at the Fermi energy that narrows as one moves towards $U_c$ from below,
and two broader incoherent features that develop at
$\pm \frac{U}{2}$, namely, the lower and upper Hubbard bands. They have
a width $\approx 2D$ and their spectral weight increases as
the transition is approached.
The insulator side, for $U > U_c$, presents only these last two high
frequency features, which are separated by an excitation gap of size
$\Delta \approx U-2D$. The different structures of the DOS (Fig.\ref{fig1})
give rise to very different optical responses.
Lets first consider the insulator, which is simpler. In this case, optical
transitions are possible from the lower to the upper Hubbard band.
We therefore expect the
optical spectrum that results from the convolution (\ref{sigma})
to display a single broad feature that extends approximately
from $U-2D$ to $U+2D$ (Fig.\ref{fig1}). A negligible temperature dependence
of the spectra is expected, as long as $T << \Delta$.
On the other hand, in the metallic case,
the low temperature optical spectrum
displays various
contributions:
{\sl i)} A narrow low frequency peak that is due
to transitions within the quasiparticle resonance, in the $T=0$ limit this
peaks becomes a $\delta-$function and is the Drude part of the optical
response.
{\sl ii)} At frequencies of order
$U\over2$ an incoherent feature of width $\sim 2D$
emerges due to
transitions between the Hubbard bands and the central resonance. {\sl iii)}
A last contribution at frequency $\sim U$ appears due to transitions
between the Hubbard bands. This is a broad feature of width $\sim 4D$.
Therefore, we expect
an optical spectrum which is schematically drawn in Fig.\ref{fig1}.
It is important to realize that, unlike the insulator, a notable
temperature dependence of the spectra is to be expected.
There is a low energy scale $T_{coh}$ that corresponds to the temperature
below which coherent quasiparticle excitations are sustained. It roughly
corresponds to the the width of the resonance
at the Fermi energy $\epsilon^*_F \equiv ZD$. As $T$ is then
increased and becomes comparable to $T_{coh}$,
the quasiparticles are destroyed, and in consequence, the contributions to
the optical spectra
associated with them, $(i)$ and $(ii)$, rapidly decrease.

It should be clear that in our previous discussion we have assumed
that the system does not order magnetically, as paramagnetic solutions were
considered. This situation can in fact  be realized by introduction of
disorder (e.g. a random distribution of $t_{ij}$) or next nearest neighbor
hopping, and avoids the artificial
nesting property of the bipartite lattice \cite{agwk,rkz}.

\subsection{Periodic Anderson Model.}

We now present a schematic discussion of
the periodic Anderson model solution.
In this case there are two different types of electrons, the $c$-electrons
which form a band
and the $d$-electrons with localized orbitals.
In the non-interacting particle-hole
symmetric case, the hybridization amplitude $V$ opens a gap
in the $c-$electron density of states  $\Delta_{ind} \sim V^2/D$.
On the other hand, the original
$\delta-$function peak of the localized $d-$electrons broadens by
hybridizing with the conduction electrons and
also opens a gap $\Delta_{ind}$.

\begin{figure}
\centerline{\epsfxsize=2.9truein
\epsffile{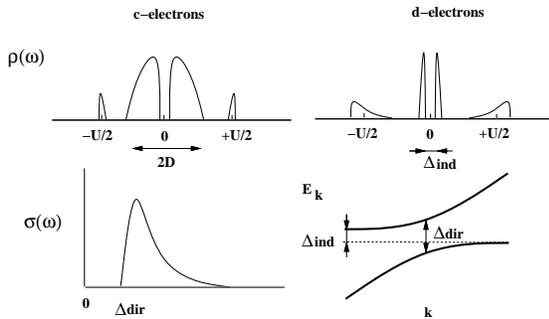}}
\caption{Schematic DOS (1/2 filling) for $c$ and $d-$electrons in the
PAM (top). The corresponding schematic optical spectra at $T=0$
(bottom left) and the schematic band structure with the direct and indirect
gaps (bottom right).
}
\label{fig2}
\end{figure}

When the effect of the interaction term is considered, as the local repulsive
$U$ is increased, one finds that for low frequencies the non-interacting
picture which was just described still holds,
however, with the bare hybridization
$V$ being renormalized to a smaller value $V^*$. Thus, we say we have a
hybridization band insulator with the hybridization amplitude renormalized
by interactions.
This can also be interpreted by considering that the interacting
$d-$electrons
form a band of ``Kondo-like'' quasiparticles, that allows to define
a coherence temperature
$T^*$ similar to the $T_{coh}$ introduced before. This coherent band
further opens a gap due to the periodicity of the lattice.
This is the well known scenario that is born out from slave boson
mean field theory and variational calculations \cite{piersrice}.
On the other hand, the present dynamical mean field theory also
captures the high energy part of the $d-$electron density of states that
develops incoherent satellite peaks at frequencies $\pm \frac{U}{2}$ with
spectral weight that is transferred from low frequencies.
Consequently,
the $c-$electron density of states is mainly made of a central broad band
of half-width $D=2t$ and a gap at the center that gets narrower as
$V \rightarrow V^*$. Also, it develops some small high frequency
structures, that result
from the hybridization with the d-electrons.
In Fig.\ref{fig2} we schematically present the density of states for
the $c$ and $d-$electrons.
As in the Hubbard model, we assume the absence of magnetic
long range order
(MLRO). For a study of the magnetic phase
of the Anderson model see Ref.\onlinecite{pamt0}.

Since the $d-$sites are localized orbitals,
only the $c-$electrons contribute to the optical response of this system.
At $T=0$, following the previous interpretation in terms of
a renormalized non-interacting hybridization band insulator
and equation (\ref{sigma}),
we expect to find an optical conductivity spectra
with a gap $\Delta_{dir}$, which decreases as the interaction is increased.
We also expect
that $\Delta_{ind} << \Delta_{dir}$,
as the first corresponds to the {\it indirect}
gap from the density of states $\Delta_{ind} \sim {V^*}^2/D$,
while the second is the {\it direct} gap $\Delta_{dir}  \sim {V^*}$
that is defined as the minimum energy for
interband transitions at a given $k$ (see Fig.\ref{fig2}).
We do not expect any other important
contributions to the optical response since, as we argued before, the
incoherent high frequency structures of the $c-$electron density of states do
not carry much spectral weight. In Fig.\ref{fig2} we schematically present the
optical response at $T=0$.

As the temperature is increased the gap in the optical
conductivity becomes gradually filled.
At high temperatures a simple picture of electrons scattering off
local moments emerge. The crossover between these two regimes, would naively
occur at
a temperature of the order of $\Delta_{ind}$.

Thus, we note that in the Hubbard model and in the periodic Anderson model the
destruction of a coherent quasiparticle state that sets the low
energy scale of the system has rather opposite effects in the optical
response. In the
first case, the correlated metallic state is destroyed as $T$ becomes of the
order of the renormalized Fermi energy, and the Drude part of the
optical response is transferred to higher energies as the insulating
state sets in.
In the second case, however, the destruction of the coherent excitations
is accompanied by the thermal
closing of the gap in the density of states that turns
the system metallic. As a consequence, the gap of the optical response
is filled with spectral weight from higher energies to become a broad
Drude-like feature.

\section{Results}
\label{res}
\subsection{Hubbard model.}

We will discuss the results for the model in regard of
different experimental data on the $V_2O_3$ system.
Vanadium oxide has three
$t_{2g}$ orbitals per $V$ atom which are filled with two electrons.
Two electrons (one per $V$) are engaged in a strong cation-cation bond,
leaving the remaining two in a twofold degenerate
$e_g$ band \cite{castellani}.
LDA calculations give a bandwidth of $\sim 0.5eV$ \cite{lda}.
The Hubbard model ignores the degeneracy of the band which is
crucial in understanding the magnetic structure \cite{castellani}, but
captures the interplay of the electron-electron
interactions and the kinetic energy.
This delicate interplay of itinerancy and localization is responsible for
many of the anomalous properties of this compound,
 and it is correctly predicted
by this simplified model.

\begin{figure}
\centerline{\epsfxsize=2.9truein
\epsffile{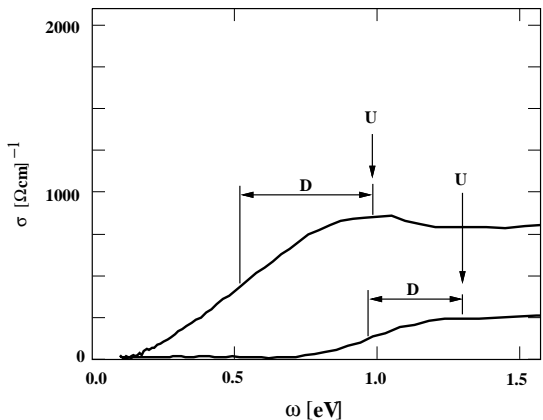}}
\caption{The experimental $\sigma(\omega)$ of insulating
$V_{2-y}O_3$ with $y=0.013$ at $10K$ (upper) and $y=0$ at $70K$ (lower).
We indicate in the spectra the position of the maxima and their width from
which the parameters $U$ and $D$ for the model calculations are extracted.
}
\label{fig3}
\end{figure}

\begin{figure}
\centerline{\epsfxsize=2.9truein
\epsffile{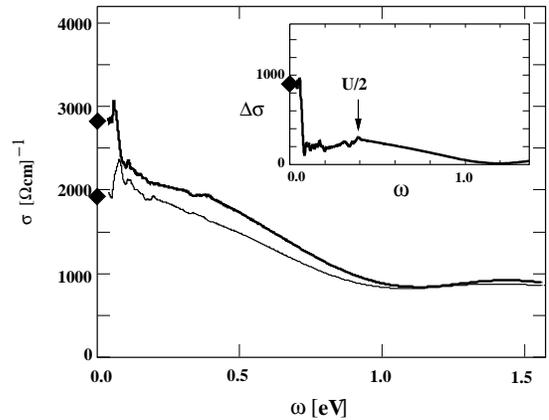}}
\caption{The experimental $\sigma(\omega)$ of metallic $V_2O_3$
at $T=170K$ (upper) and $T=300K$ (lower).
The inset contains the difference
of the two spectra
$\Delta\sigma(\omega)=\sigma_{170K}(\omega)-\sigma_{300K}(\omega)$.
Diamonds indicate the measured dc conductivity $\sigma_{dc}$.
}
\label{fig4}
\end{figure}

Experimentally one can vary the parameters $U$ and $D$, by introducing $O$ and
$V$ vacancies or by applying presure or chemical substitution of
the cation.
We can use experimental data to extract approximate parameters to
be used as input to our model. In particular,
from the experimental optical conductivity
data in the insulating phase, a rather accurate determination
can be made because, as it is apparent from the spectra, the
low frequency contribution is mainly due to a single peak \cite{thomas}.
In regard of our schematic discussion of the previous section,
the position
of the maximum should approximately correspond to the parameter
$U$ that corresponds to transitions from the lower to the upper
Hubbard band. Also,
according to the picture of the previous section,
the total width of the peak should be $\sim 4D$ which is twice the width
of the Hubbard bands. Therefore, we can approximately estimate
the parameter $D$ as the distance from position of the peak maximum
to the frequency where the feature decreased to half its height (see
Fig.\ref{fig3}).
The parameters from the metallic optical conductivity
spectra are not so easily extracted. However, we can still obtain a rather
precise determination by considering the {\it difference spectra}
between the
data at $170K$ and at $300K$ (see inset of Fig.\ref{fig4}).
As we shall later discuss in detail, it turns out that the
feature that appears in the difference spectra at a frequency
$\approx 0.4eV$ can be associated with the parameter $U/2$. This
is also intuitively suggested by the schematic discussion of the previous
section, as this feature corresponds to the enhancement of transitions
from the lower Hubbard band to the central resonance at the Fermi level
and from the resonance to the upper Hubbard band which are at a distance
$\sim U/2$. The value for the parameter $D \approx 0.4eV$
in the metallic phase
was determined by noting that: i) {\it a priori} there is no reason to expect
that it should be much different than in the insulating phase (unlike
the parameter $U$ which could be modified by screening);
ii) it is consistent with the recent LDA calculation that
gives a half-width of $\approx 0.5eV$ for the narrow bands
at the Fermi level \cite{lda}; iii) despite the lack of very good
experimental resolution the value is consistent with both
the optical data that we reproduce in Fig.\ref{fig4}
and photoemission experiments \cite{photo};
iv) as will be shown later in the paper, this estimated value will
allow to gather in a single semi-quantitative consistent picture
the optical conductivity results with the $V_2O_3$ phase diagram
and experimental results for the slope of the specific heat.
The extracted parameters along with the values for the size of the
optical gap (in the insulators) and the total optical
spectral weight are summarized in table \ref{table1}.
\begin{table}
\begin{center}
\begin{tabular}{lcccc}
Phase & \multicolumn{4}{c}{Parameter }  \\
  &D [eV] &U [eV] &$\Delta$ [eV] & $\protect \omega_P^2/4\pi$ [eV/$\Omega$cm]
 \\
 \tableline
Ins. (y=0)  & $.33 \pm .05$ & $1.3 \pm .05$ & $.64 \pm .05$ &
$170 \pm 20$  \\
Ins. (y=.013) & $.46 \pm .05$ & $.98 \pm .05$ & $.08 \pm .05$ &
$800 \pm 50$ \\
Metal (170K)  &$ .4  \pm .1$ &$ .8 \pm .1$ & -- &
$ 1700 \pm 300$ \\
\end{tabular}
\caption{Experimental parameters for the model.}
\label{table1}
\end{center}
\end{table}

\begin{figure}
\centerline{\epsfxsize=2.9truein
\epsffile{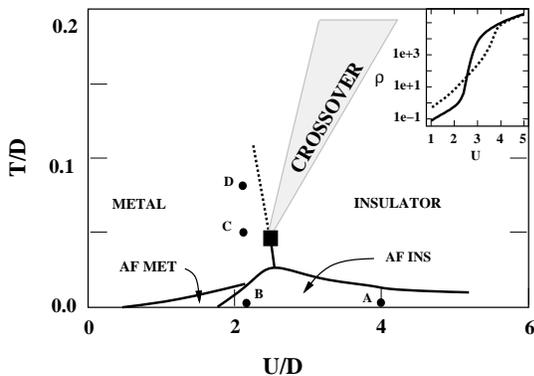}}
\caption{Approximate phase diagram for the model with $n.n.$ and
$n.n.n.$ hopping
$(t_2/t_1)=\protect\sqrt{1/3}$.
The $1st$ order paramagnetic metal-insulator transition ends at the critical
point $T_{MIT}$ (square).
The dotted line and the shaded region describe two crossovers as  discussed
in the text.
The full circles indicate the position
of the optical spectra. $A$: insulator
($y=0$), $B$: insulator ($y=.013$), $C$: metal ($y=0,\ 170K$),
$D$: metal ($y=0,\ 300K$).
Note that for comparison with experimental results increasing $U/D$
is associated with decreasing pressure.\protect\cite{kuwamoto,mcwhan}
Inset: $\rho_{dc}(U)$ for $T=0.06D$ (full) and $T=0.15D$ (dotted).
}
\label{fig5}
\end{figure}

\begin{figure}
\centerline{\epsfxsize=2.9truein
\epsffile{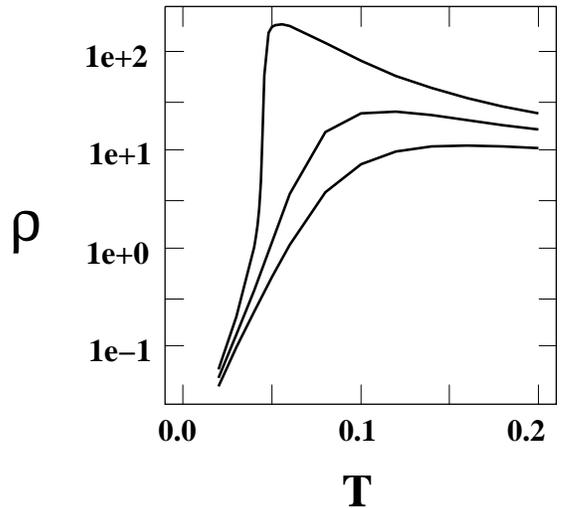}}
\caption{
$\rho_{dc}(T)$ for $U/D=2.1,2.3,2.5$ (bottom to top).
The maxima of $\rho_{dc}(T)$ defines the dotted line.
Obtained with the IPT method.
}
\label{fig5a}
\end{figure}

In Fig.\ref{fig5} we display the phase diagram of
the model in large dimensions generalized to include $n.n.$ hopping $t_1$ and
$n.n.n.$ hopping $t_2$. The condition $t_1^2 + t_2^2 = t^2$ keeps the
bare density of states $\rho^0$ invariant. For $t_2/t_1 =0$ we recover the
original hamiltonian, and $t_2/t_1 =1$ gives the paramagnetic solution.
The extra hopping provides a magnetically frustrating interaction \cite{rkz}.
This phase diagram obtained for $t_2/t_1 =\sqrt{1/3}$ has the same topology
as the experimental one \cite{kuwamoto,mcwhan,carter}.
Frustration lowers the $T_{Neel}$ well below  the second order
 $T_{MIT}$ point \cite{rkz}.
Using the parameters of table \ref{table1},
$T_{MIT}\approx 240K$,
which is only within less than a factor of 2 from the experimental result.
The dotted line  indicates a crossover separating
a good metal at low $T$ and a semiconductor  at higher $T$.
Between these states
$\rho_{dc}(T)$ has an anomalous rapid increase, as is shown in the
results of Fig.\ref{fig5a}.
The reason for this feature
can be traced to the thermal destruction of the coherent central quasiparticle
peak in the $DOS$.  We find the behavior of $\rho_{dc}(T)$ to be in good
agreement with the
experimental results of Mc Whan {\sl et al.} \cite{mcwhan}.
Another crossover is indicated by a shaded area, it
separates a semiconducting region with a gap
$\Delta$ comparable with $T$ from a good insulator where $T << \Delta$,
consequently the crossover temperature  increases  linearly
with $U$ and the horizontal width of the crossover region
becomes broader with increasing $T$.
This crossover is characterized
by a sudden increase in $\rho_{dc}$ as function of
$U$ at a fix $T$ (inset Fig.\ref{fig5}).
This crossover behavior was experimentally
observed in $V_2O_3$ by Kuwamoto {\sl et al.} \cite{kuwamoto}.

\subsubsection{Magnetically ordered solutions.}
In infinite dimensions  the
optical conductivity
is a weighted  convolution  of two
one particle  spectral functions.
The one particle spectral function is, therefore, the basic building
block which gives rise to the various features of the optical conductivity.
In this subsection we consider
the nature of the spectral functions with magnetic long range order (MLRO).
The understanding of the qualitative differences and similarities between
solutions with and without MLRO is relevant in regard of systems,
like $V_2O_3$, that present both
antiferromagnetic (AFI) and paramagnetic (PI)
insulating phases.
Results for the corresponding optical spectra will follow in the next section.

\begin{figure}
\centerline{\epsfxsize=2.9truein
\epsffile{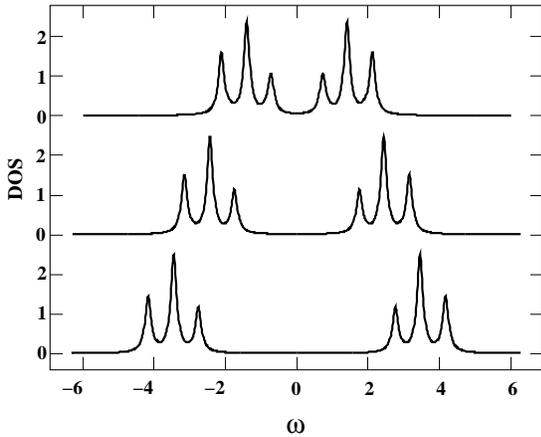}}
\caption{DOS of the paramagnetic insulator solution
obtained from exact diagonalization of 7 sites
with $U=3,5,7$ (top to bottom).
A small broadening has been added to the poles.
}
\label{fig5b}
\end{figure}

\begin{figure}
\centerline{\epsfxsize=2.9truein
\epsffile{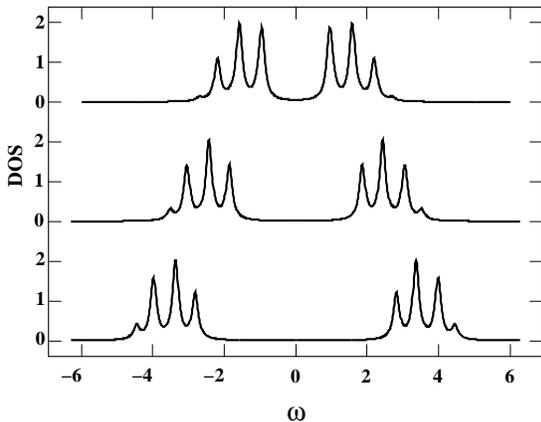}}
\caption{DOS of the antiferromagnetic insulator solution
obtained from exact diagonalization of 7 sites
with $U=3,5,7$ (top to bottom).
A small broadening has been added to the poles.
}
\label{fig5c}
\end{figure}

In Figs \ref{fig5b} and \ref{fig5c} we respectively show the single particle
spectra of the PI and AFI insulating solutions for different values of
the interaction $U$.
The results are obtained from the ED method at $T=0$ for clusters of
seven sites. The finite number of poles in the spectra correspond
to the finite size of the clusters that can be practically considered.
A finite broadening of the poles was added for better visualization.

In the AFI case, we plot the averaged value of the sublattice Green functions
$\bar G_\sigma$ \cite{gotoref}
\begin{equation}
\bar G_\sigma={1\over 2}(G_{A\sigma}+G_{B\sigma})=
{1\over 2}(G_{A\sigma}+G_{A\ -\sigma})
\end{equation}
which is the quantity to be compared to photoemission experiments.

It is interesting to realize from these results, which correspond to rather
large values of the interaction $U$, that the spectra in both cases are
roughly similar. They both present a lower and upper Hubbard bands at energies
$\approx \pm {U\over 2}$ with a bandwidth $\approx 2D$ and a corresponding
gap $\Delta \approx U-2D$.

In particular, the PI solution, merely presents a rigid shift of the
incoherent Hubbard
bands as the interaction $U$ is varied, which is reminiscent of
Hubbard's solution to the model \cite{zrk,hub3}.
On the other hand, in the AFI case, the shape of the density of states
follows from the fact that the sublattice magnetization is basically saturated
at these large values of the interaction.

At $U=7$, the largest value of the interactions considered,
we observe
that the shape of the spectra of $\bar G$ becomes very similar to the
corresponding one in the disordered case. This can be understood from the
fact that the magnetic exchange scale $J \sim {D^2 \over U}$ vanishes as
$U$ becomes large.
As one decreases the strength of the interaction, we observe that the AFI
spectra
become increasingly different from the PI ones. In the former
there is a transfer
of spectral weight that occurs within the bands, from the higher to the
lower frequencies. This is a consequence of the fact that, as the scale $J$
becomes increasingly relevant, the spectra acquire a more coherent character.
The ``piling up'' that occurs with the transfer of spectral
weight as $U$ is reduced, is the precursor of the weak coupling
inverse square root  singularity in the low frequency part of the density of
states.
It is interesting to note that recent photoemission experiments in
$V_2O_3$ report the presence of a small anomalous enhancement in the lower
frequency edge of the spectrum in the AFI phase. This
feature may be interpreted from the previous results as evidence of the
transfer of weight within the Hubbard bands.

\begin{figure}
\centerline{\epsfxsize=2.9truein
\epsffile{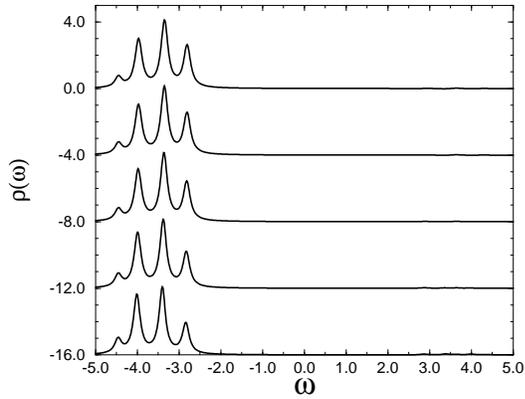}}
\caption{$\epsilon$-resolved single particle spectra
of the antiferromagnetic insulator solution
obtained from exact diagonalization of 7 sites
with $U=7$ with $\epsilon=0.0,0.25,0.5,0.75,1.0$ (top to bottom).
A small broadening $\eta=0.1$ was added to the poles and the figures were
vertically shifted for better visualization.
}
\label{fig5d}
\end{figure}

\begin{figure}
\centerline{\epsfxsize=2.9truein
\epsffile{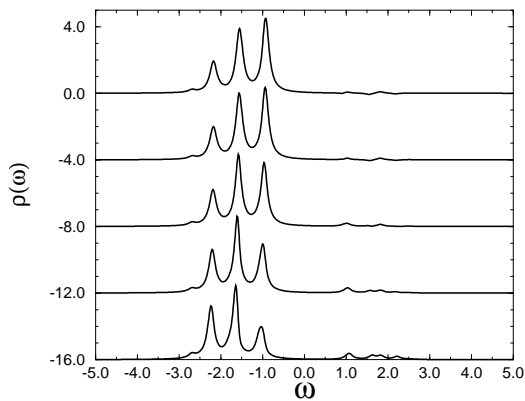}}
\caption{$\epsilon$-resolved single particle spectra
of the antiferromagnetic insulator solution
obtained from exact diagonalization of 7 sites
with $U=3$ with $\epsilon=0.0,0.25,0.5,0.75,1.0$ (top to bottom).
A small broadening $\eta=0.1$ was added to the poles and the figures were
vertically shifted for better visualization.
}
\label{fig5e}
\end{figure}

A complementary perspective on the results that we just discussed is
obtained by looking at the $k$-resolved spectra given by the
imaginary part of the Green function $G(k,\omega)$ which reads,
\begin{equation}
G(k,\omega)=\frac{1}{\omega-\epsilon_k -\Sigma(\omega)}.
\end{equation}
In the large $d$ limit
the
Green functions are labeled by the energy $\epsilon$ \cite{vollmetz}.
Nevertheless, one can still think of this quantity
as the analogous of the $k-$resolved spectra if one notes that
the $\epsilon$ goes from $-D$ to $D$ as it traverses the band (the dispersion
is linear in the
non interacting case). Thus, we can associate the ``zone center vector''
$\epsilon = 0$ with the
$\Gamma$ point of the BZ, and the ``nesting vectors''
$\epsilon = \pm D$ with the commensurate $M$ point.
In Figs.\ref{fig5d} and \ref{fig5e}
we show the $\epsilon$-resolved spectra for the different
values of the interaction considered before. From the inspection
of the spectra we observe that in the low $J$ case for $U=7$ the
single particle spectra remain basically unmodified as we scan
the $\epsilon$ ``wave vectors'', which indicates the very incoherent
character of the single particle excitations.
On the other hand, as we lower $U$ and the scale $J$ becomes larger,
we note the emergence of a peak in the $U=3$ case for small values of
$\epsilon$. Notice, also, the more dispersive character
of the excitations.

\begin{figure}
\centerline{\epsfxsize=2.9truein
\epsffile{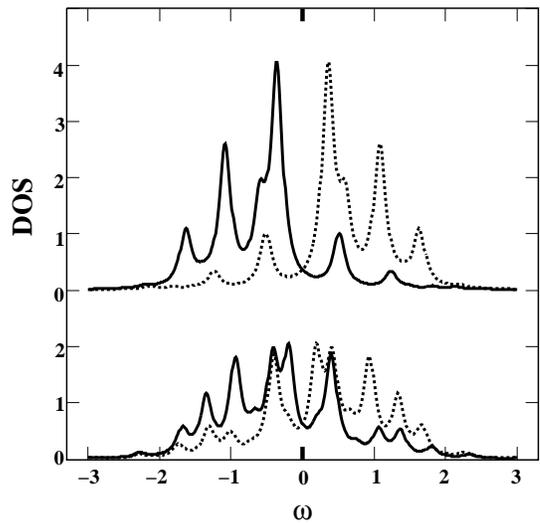}}
\caption{DOS obtained from exact diagonalization of 7 sites
with $U=1.5$ for $(t_2/t_1)=0$ (top) and $(t_2/t_1)=\protect\sqrt{1/3}$
(bottom). A small broadening $\eta=0.1$ was added to the poles.
}
\label{fig6}
\end{figure}

An important parameter of the theory is the degree of magnetic frustration.
Frustration is necessary not only to obtain
the observed phase diagram of $V_2 O_3$ \cite{rzk} but to account for the
general shape of its angular integrated photoemission spectra \cite{shin}.
We can summarize the results of this section by saying that the
ED solutions indicate that as the degree of frustration is reduced
and as $U/t$
is reduced, the spectral function develops more dispersion, and the
excitations at low energy are more coherent (i.e. the imaginary part
of the self energy is smaller).
Many experiments place $V_2 O_3$ in the regime of strong frustration, while
the observation of dispersive features in the insulating phase of
$Ni S_{1.5} Se_{0.5}$ \cite{shen}
may be explained by a lower degree of magnetic frustration
in this compound.
We shall now briefly consider the
antiferromagnetic metallic state (AFM) that
occurs upon the introduction of next
nearest neighbor hopping in the nested lattice (i.e. partial frustration)
at small (but finite) values of $U$ (see phase diagram in Fig.\ref{fig5}).
We obtain the density of states
for $t_2/t_1 = 0$ (AFI), and $t_2/t_1 =
\sqrt{1/3}$ (AFM) with the interaction
$U= 1.5$. The results obtained from 7 sites exact diagonalization
are shown in Fig.\ref{fig6}.
It is very interesting to note that the peak structure of the density of
states seems to be
divided into low frequency features near $\omega = 0$, and higher frequency
structures at energies of the order of the band-width (which is also
comparable to $U$ for the chosen parameters). This is even more clear
in the antiferromagnetic metallic state  with partial frustration.

\begin{figure}
\centerline{\epsfxsize=2.9truein
\epsffile{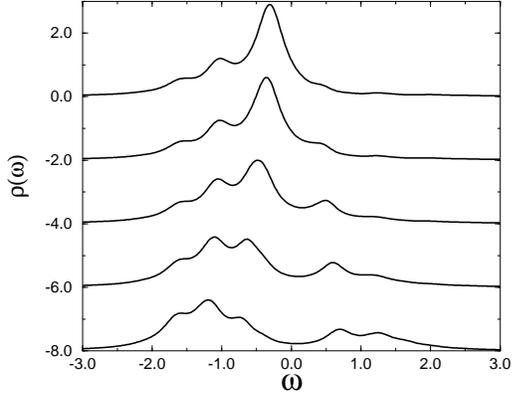}}
\caption{$\epsilon$-resolved single particle spectra
of the antiferromagnetic insulator solution
obtained from exact diagonalization of 7 sites
with $U=1.5$ with $\epsilon=0.0,0.25,0.5,0.75,1.0$ (top to bottom).
A small broadening $\eta=0.2$ was added to the poles
and the figures were
vertically shifted for better visualization.
}
\label{fig6a}
\end{figure}

We note that our
results are qualitatively similar to the recent exact diagonalization
results for the $t-J$ model
and also to
quantum Monte Carlo results for the Hubbard model on
2-dimensional finite size lattices
with a choice of parameters comparable to the one used here \cite{moreo}.
Furthemore, the similarity to the physics found in
finite dimensional finite size lattices becomes more striking
by comparing the $\epsilon$-resolved spectra of Fig.\ref{fig6a}
with the ones obtained by Preuss {\it et al.} in a recent
QMC study \cite{preuss}.

An important technical
remark is that in order to apply the exact
diagonalization method of Ref.\onlinecite{srkr}
to the problem with intermediate
frustration $0 < t_2/t_1 < 1$, it is necessary to be able to
average the continued fractions for the
spin-up and spin-down Green functions into a single
continued fraction.
To perform this task
we use the algorithm  detailed in the Appendix.

\subsubsection{The insulating state.}

\begin{figure}
\centerline{\epsfxsize=2.9truein
\epsffile{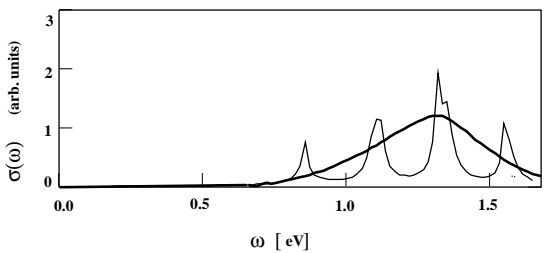}}
\caption{The model $\sigma(\omega)$ for the insulating
solution results at $U=4D$ and $T=0$ from ED (thin) and IPT (bold).
}
\label{fig7}
\end{figure}

We now turn to the optical conductivity results.
The experimental optical
spectrum of the  insulator was
reproduced in Fig.\ref{fig3} \cite{thomas2}.
It is characterized by an excitation gap at
low energies, followed by an incoherent feature that corresponds to
charge excitations of mainly Vanadium character \cite{thomas}.
These data are to be compared with the model results of
Fig.\ref{fig7}.
The overall shape of the
spectrum is found to be in very good agreement with the
experimental results for the pure $V_2O_3$ sample.
We display the optical spectra results from both IPT and
the ED method. The data show the very good agreement between
this two methods. The peak structure in the ED data is due to the
finite number of poles that result from the
finite size of the clusters that can be considered in practice.

\begin{figure}
\centerline{\epsfxsize=2.9truein
\epsffile{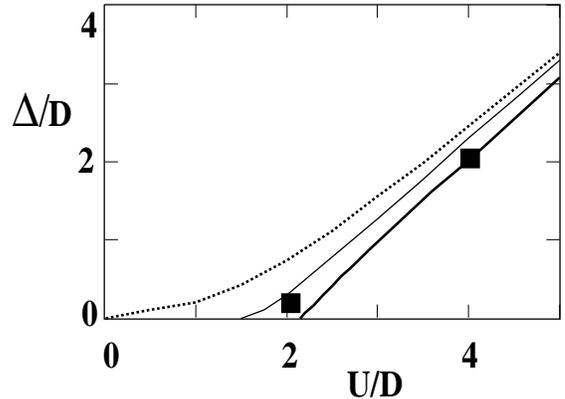}}
\caption{The gap
$\Delta$ versus $U$ for
the antiferromagnetic, partially frustrated and paramagnetic
insulators (dotted, thin and bold). $\Delta$ is twice the energy of the
lowest pole from the ED Green function.
The data are for $n_s \rightarrow \infty$
from clusters of 3, 5 and 7 sites assuming ${1}/{n_s}$ scaling behavior.
Black squares show the experimental gap for $V_{2-y}O_3$ with $y=0.0$ and
$0.013$.
}
\label{fig7a}
\end{figure}

In Fig.\ref{fig7a} we display the results
for the size of the gaps $\Delta$, which
are in excellent agreement with the experimental results
indicated by black squares \cite{thomas2}.
It is interesting to note that the results of Fig.\ref{fig7a},
shown for various degrees of magnetic frustration,
indicate that in $V_2O_3$ frustration plays an important role.
The experimental
system seems to be closer to the limit of strong frustration, which is
consistent with neutron scattering results that indicate different signs for
the magnetic interactions between different neighboring sites \cite{neutron}.

\begin{figure}
\centerline{\epsfxsize=2.9truein
\epsffile{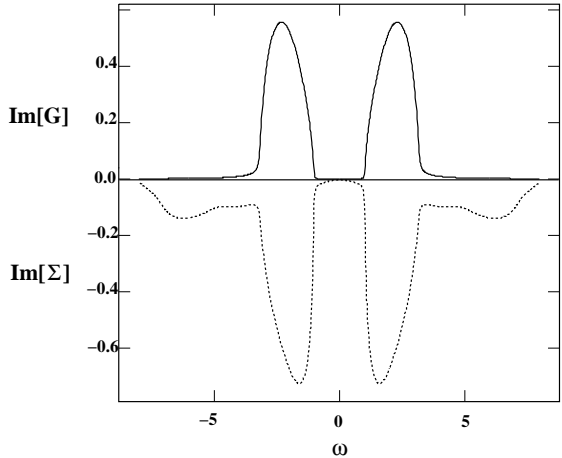}}
\caption{${\rm Im}[G(\omega)]$ and ${\rm Im}[\Sigma(\omega)]$ for $U=4$ from
IPT. Note that ${\rm Im}[\Sigma(\omega)]$ is large when
${\rm Im}[G(\omega)]$ is non-zero
indicating the incoherent character of the particle excitations.}
\label{fig8}
\end{figure}

Another interesting point is
the fact that the gap obtained in the model optical spectra
and the one obtained from the position of the poles in the
single particle spectra coincide (Figs.\ref{fig7} and \ref{fig7a}).
We therefore conclude that in this model the direct and indirect
gap are very close (which justifies a posteriori that $\Delta$ is measured
from the lowest pole of the local Green function).
This result already predicted in Ref. \onlinecite{opt},
was experimentally confirmed by accurate recent photoemission
study of $V_2O_3$ \cite{shin}.
This follows from
the fact that the imaginary part of
the self-energy is very large wherever the electron density of states
is non zero in the insulating solution (see Fig.\ref{fig8}).
This is nothing but a direct
consequence of the complete incoherent character of the upper and lower
Hubbard bands.
 They describe a completely incoherent propagation, and
one should not think of them as usual metallic bands ``shifted'' by the
interaction $U$ .
Notice that from  the
discussion in the previous section in the unfrustrated case,
one expects a a larger difference between the direct and the indirect gap.

\begin{figure}
\centerline{\epsfxsize=2.9truein
\epsffile{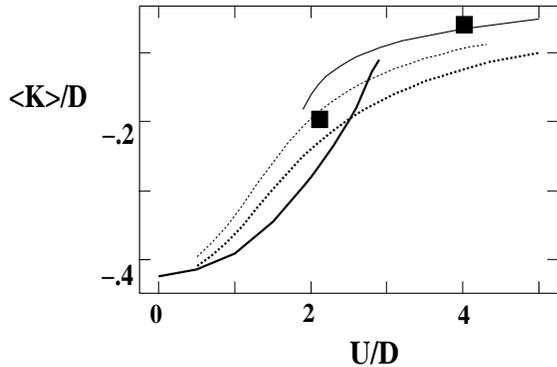}}
\caption{
Kinetic energy $\langle K \rangle$ versus
$U$ at $T=0$ for the antiferromagnetic and paramagnetic insulators
(bold-dotted and thin), paramagnetic metal (bold), and
partially frustrated model (thin-dotted).
Black squares show the insulator experimental results.
}
\label{fig9}
\end{figure}

A final and important quantity that can be compared to the experiment
is the integrated spectral
weight $\frac{\omega_P^2}{4\pi}$ which
is related to $\langle K \rangle$ by the sum rule (\ref{sum}).
Setting the lattice constant $a\approx 3 \AA$ the average $V-V$ distance,
we find our results also in good agreement with the experiment
(see Fig.\ref{fig9}).

An interesting question, not yet  fully settled, is
the mechanism by which the insulating solution is
destroyed.
The destruction of the  insulating state
occurs at a point $U_{c1}$ which may be different from
the critical point $U_c \approx 3D$
that is associated to the breakdown of the metallic state as the interaction
$U$ is increased \cite{agwk,rkz}.
This issue is physically  relevant  because one can envision a situation
where the magnetic order stabilizes the insulating solution over the metallic
solution but due to a large degree of magnetic frustration, the insulating
solution is very close to the fully frustrated paramagnetic insulator.
The destruction of the paramagnetic insulating state was discussed
in Ref.\onlinecite{rzk} using IPT. Here we address this issue using
exact diagonalization.

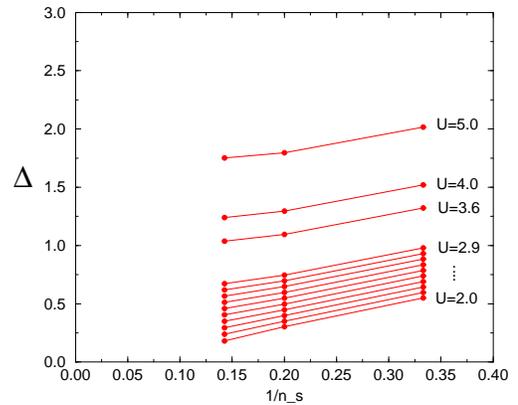
\begin{figure}
\centerline{\epsfxsize=2.9truein
\epsffile{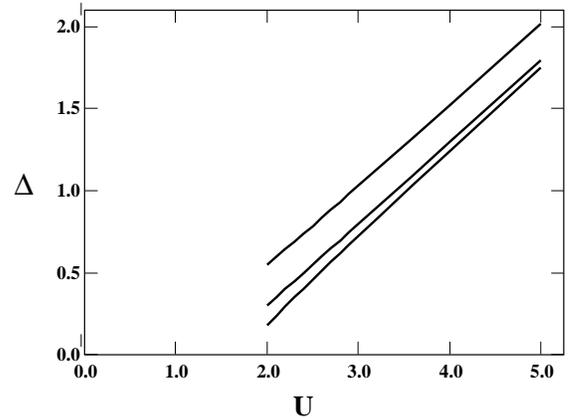}}
\caption{The gap
$\Delta$ versus the interaction $U$ in
the paramagnetic insulator. $\Delta$ is twice the energy of the
lowest pole from the ED Green function.
The data are
from clusters of 3, 5 and 7 sites (top to bottom).
}
\label{fig7b}
\end{figure}

We first study the  behavior of the  gap in the one particle
excitation spectrum defined as the position of the lowest energy
pole (with non negligible weight) in the Green function
as a function of the number of sites
included in the representation of the effective bath.
Although the mean field theory is strictly formulated in the thermodynamic
limit, in practice, the representation of the bath by a finite
number of orbitals introduces finite size effects.
The data shown
in Fig.\ref{fig7a} were obtained from the extrapolation of results from
finite size cluster Hamiltonians $H^{n_s}$ to
the $n_s \rightarrow \infty$ system.
The value for $\Delta$ is defined as twice the energy of the lowest frequency
pole appearing in the Green function.
In Fig.\ref{fig7b} we show the gap as a function
of the interaction $U$ in systems of
$n_s=3,\ 5$ and $7$ sites. Fig.\ref{fig7c} contains similar results
as a function
of $1/{n_s}$ which shows the good scaling of $\Delta$, especially as the
gap goes to zero as $U$ is decreased.
Thus, this approach indicates a continuous closure of the gap at
a critical value of the interaction $U_{c1}=2.15$.

\begin{figure}
\centerline{\epsfxsize=2.9truein
\epsffile{fig18}}
\caption{The gap
$\Delta$ versus the inverse of the number of sites $1/{n_s}$ in
the paramagnetic insulator for various values of $U$.
$\Delta$ is twice the energy of the
lowest pole from the ED Green function.
}
\label{fig7c}
\end{figure}

We also investigate the behavior of the inverse moments of
the spectral function defined as:
\begin {equation}
m_{-n} =  \int_{0}^{\infty}  \frac{\rho(\epsilon) d\epsilon }
{\epsilon^n}
\end{equation}

The behavior of these quantities give a more detailed
picture of the transition.
The local picture of the paramagnetic insulator is that of a spin embedded
in an insulator. Hybridization with the bands of this
insulator transfers spectral weight
to high frequencies but  the spin remains well defined at low energies
(even though with a reduced spectral weight) as  long as there is a
finite gap in the insulator. As $U_{c1} $ is approached, and the gap
decreases we face the question whether the spin remains well defined even
at the transition point.
This depends on the behavior of the density of states of the bath $\rho_{bath}$
at low frequecies (we recall, $\rho_{bath}$
is essentially $\rho$ in a Bethe lattice, {\it cf.} Eq.\ref{s0}).
Whittoff and Fradkin \cite{wf} have shown that
if the density of states of the bath vanishes as a power-law
$\rho_{bath} \propto \epsilon^\beta$
the spin remains well defined if $ \beta > 1$
while the spin is Kondo  quenched if $ \beta  <  1$ and the spin
degree of freedom is absorbed by the conduction electrons.
The case  $ \beta =  1$ is marginal.

In a previous publication \cite{rkz} we showed that
within IPT the second inverse  moment
remains  finite at the transition, while it diverges in
the  Hubbard III solution.

\begin{figure}
\centerline{\epsfxsize=2.9truein
\epsffile{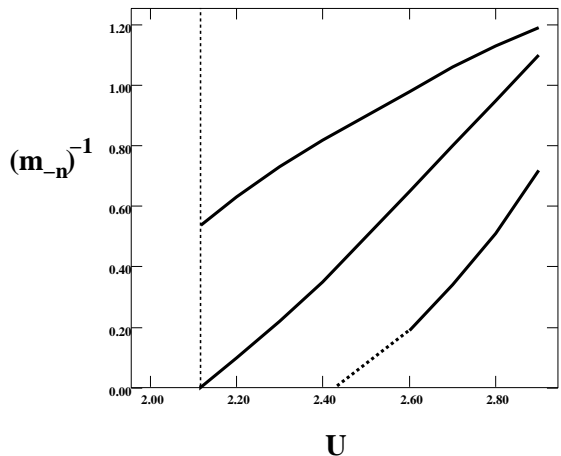}}
\caption{
Inverse of the first three inverse moments $(m_{-n})^{-1}$
of the density of states as a function of $U$.
The three curves correspond from top to bottom to the  inverse of the
first, second and third inverse moment respectively.
The results are the $n_s \rightarrow \infty$ extrapolation
from clusters of 3, 5 and 7 sites assuming ${1}/{n_s}$ scaling behavior.
The dotted continuation of the last curve indicates results where
the scaling is not
reliable due to the strong divergent behavior of that inverse moment.
The vertical dotted line indicates the value obtained for $U_{c1}$
from the inverse moment analysis.
}
\label{fig7d}
\end{figure}

Notice that  $ m_{-2}$ can remain finite up to
the transition even when the gap closes, but
a divergent second
inverse moment {\it implies} the continuous closure of the gap.
In Fig. \ref{fig7d} we plot the {\it inverse} of $ m_{-2}$
together with that of the first and third inverse moments.
The results correspond to the extrapolation to the infinite size
effective bath, performed similarly as was done previously for the gap.
The inverse of the second inverse moment  shows good scaling
behavior with the system size and is found to go to zero for
$U \approx 2.12$. At this value of the interaction
the moment diverges, which signals the breakdown of the insulating
state, with the gap closing continuously.
As expected, the first inverse moment remains finite at the
transition (it also shows good scaling behavior) and, on the other
hand, the inverse of the third inverse moment becomes negative
even before the transition. This is due to the fast divergence of the third
moment which renders
the finite size scaling inaccurate.
It is important to stress that this way of looking at the
transition is very different from the previous one, nevertheless,
the estimates for $U_{c1}$ that are obtained after the infinite size
bath extrapolation are consistently predicted to within less than 2\%.
The results are substantially different from the ones obtained from
IPT.

\subsubsection{The metallic state.}

We now  discuss the data in  the  metallic phase.
In Fig.\ref{fig4} at
the beginning of this section, we reproduced experimental data
for pure samples that become insulating below $T_c \approx 150K$
\cite{opt}.
The spectra were obtained in the metallic phase at $T =170K$ and $T=300K$ and
are made up of broad absorption at higher frequencies
and some phonon lines in the far infrared.
They appear to be rather
featureless, however, upon considering their difference
(in which the phonons are approximately eliminated)
distinct features are observed.
As $T$ is lowered, there is
an enhancement of the spectrum at intermediate frequencies of
order $0.5eV$; and more notably, a sharp low
frequency feature emerges that extends from $0$ to $0.15eV$.
Moreover, these enhancements result in an anomalous {\it change} of
the total spectral weight $\frac{\omega_P^2}{4\pi}$ with $T$.
We argue below, that these observations can be accounted by
the Hubbard model treated in mean field theory.

\begin{figure}
\centerline{\epsfxsize=2.9truein
\epsffile{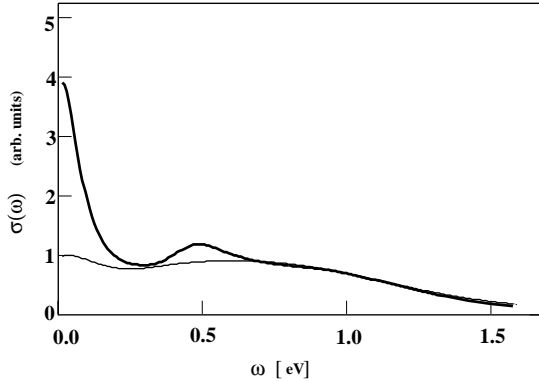}}
\caption{The model $\sigma(\omega)$ for the metallic solution
at $U=2.1D$ and $T=0.05D$ (upper) and $0.083D$ (lower).
A small $\Gamma = 0.3$ and $0.5D$ was included to
mimic a finite amount of
disorder.
}
\label{fig10}
\end{figure}

In Fig.\ref{fig10} we show the calculated  optical spectra
obtained from IPT for two different values of $T$.
The interaction is set to $U=2.1D$ that places the system
in the correlated metallic state.
It is clear that,
at least, the qualitative aspect of
the physics is already captured
and setting $D \approx 0.4eV$ we find these results
consistent with the experimental data on $V_2O_3$ (Fig.\ref{fig4}).
As the temperature is lowered, we
observe the enhancement of the incoherent structures at
intermediate frequencies of the order $\frac{U}{2}$ to $U$
and the rapid emergence
of a feature at the lower end of the spectrum.
This two emerging features can be interpreted from the qualitative
picture that was discussed in Sec.\ref{phys}
which is relevant for low $T$.
{}From the model calculations with the parameters of table I
we find the enhancement of the spectral weight
taking place
at a scale $T_{coh} \approx 0.05D \approx 240K$
which correlates well with the experimental data.
$T_{coh}$ has the physical meaning of the
temperature below which the Fermi
liquid description applies,\cite{rkz}
as the quasiparticle resonance emerged in the density of states.

\begin{figure}
\centerline{\epsfxsize=2.9truein
\epsffile{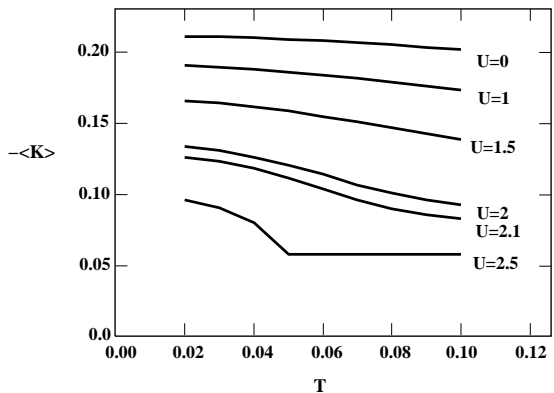}}
\caption{Expectation value of minus the kinetic energy $\langle K \rangle$
as a function of the temperature for various $U$ (IPT).
This quantity is directly proportional to the optical conductivity
sum rule. It predicts a notable {\it increase} in the optical
spectral weight as the temperature is {\it decreased} in the
correlated metallic regime.}
\label{fig11}
\end{figure}

In Fig.\ref{fig11} we present the
results for $\langle K \rangle$ as a function
of the temperature.
An interesting prediction of the model is the anomalous increase
of the integrated spectral
weight $\frac{\omega_P^2}{4\pi}$ as $T$ is decreased, a feature that
is actually observed in the experimental data (note that the spectral
weight is {\it not} recovered upto  the highest frequencies where
experimental data is available ($\omega \approx 6eV$).
This effect is due to the rather strong $T$ dependence of the kinetic
energy $\langle K \rangle \propto \frac{\omega_P^2}{4\pi}$
in the region near the crossover indicated by a dotted
line in the phase diagram (Fig.\ref{fig5}).
It results from the competition between two small energy scales, namely,
the temperature and the renormalized Fermi energy $\epsilon_F^*$.

\begin{figure}
\centerline{\epsfxsize=2.9truein
\epsffile{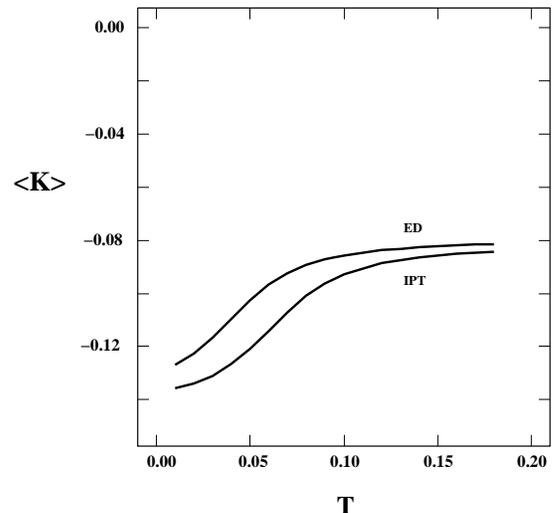}}
\caption{Comparison of the
expectation value of the kinetic energy $\langle K \rangle$
as a function of the temperature for $U=2$ as obtained from IPT and ED method.
}
\label{fig11a}
\end{figure}

Fig.\ref{fig11a} contains the comparison between the
results for the same quantity $\langle K \rangle$
at $U=2$ as obtained from the IPT
and the finite temperature ED method. It demonstrates that the temperature
dependence is indeed a true feature of the model which is being successfully
captured by the approximate IPT solution.

Although the qualitative aspect seems to be very accurately described
by the model,
we find $\frac{\omega_P^2}{4\pi} \approx
1000 \frac{ev}{\Omega cm}$ which is somewhat lower than the experimental
result.
This could be due to the contribution
from tails of bands at higher energies that are not
included in our model, or it may indicate that the
bands near the Fermi level are degenerate.

We now want to finally consider an important prediction of the model for the
slope of the linear term in the
specific heat $\gamma$  in the metallic phase.
Experiments show that  the slope
$\gamma$  is in general unusually  large.
For $0.08\  Ti$ substitution $\gamma \approx 40 \frac{mJ}{mol K^2}$, while
for a pressure of $25 Kbar$ in the pure compound
$\gamma \approx 30 \frac{mJ}{mol K^2}$ and  with $V$ deficiency in
a range of $y=0.013$ to $0.033$ the value is
$\gamma \approx 47 \frac{mJ}{mol K^2}$.\cite{gamma}
In our model $\gamma$ is simply related to the weight in
the Drude peak in the optical conductivity and to the quasiparticle
residue $Z$ ,
$\gamma = \frac{1}{ZD}3 \frac{mJ eV}{mol K^2} $.
The values of  $U=2.1D$ and
$D \approx 0.4eV$
 extracted from the optical data
correspond to a quasiparticle residue $Z \approx 0.3$,
and result in $\gamma \approx  25  \frac{mJ}{mol K^2}$ which
is close to
the experimental findings.
Thus, it turns out that the mean field
theory of the Mott transition
explains in a natural and qualitative manner, the experimentally observed
optical conductivity spectrum,  the
anomalously large values of the slope of the specific heat $\gamma$,
and the dc-conductivity in the metallic phase,
as consequence of the appearance of a single small energy scale,
the renormalized Fermi energy $\epsilon^*_F$.

\subsection{Periodic Anderson model.}

\subsubsection{Gap formation.}
A second class of
systems where the correlations
induce an
 anomalous temperature dependence are the Kondo insulators. While the
most qualitative physics of these systems is well understood, several features
remain puzzling \cite{aepplifisk}.
The charge gap $\Delta_c$ measured in optical conductivity
is larger than the spin gap $\Delta_s$ measured in neutron
scattering \cite{bucher}. The
transport gap $\Delta_t$ obtained from the activation energy in dc-transport
measurements is smaller than $\Delta_c$. The gap $\Delta_c$ begins to open at a
characteristic temperature $T^* \sim \frac{\Delta_c}{5}$ and becomes fully
developed at a much smaller temperature of the order of $\frac{T^*}{5}$. Also,
the gap is temperature independent below $T^*$.
In $Ce_3Bi_4Pt_3$ it is found that $\Delta_c \approx 450K$, $\Delta_s
\approx 250K$, $T^* \approx 100K$ and the optical gap is completely
depleted only below $\approx 25K$ \cite{bucher}. On the other hand,
qualitatively similar results were reported for $FeSi$, with
$\Delta_c \approx 1000K$, $T^* \approx 200K$ and the gap
becomes depleted between $20$ and $100K$ \cite{schlesinger}.

\begin{figure}
\centerline{\epsfxsize=2.9truein
\epsffile{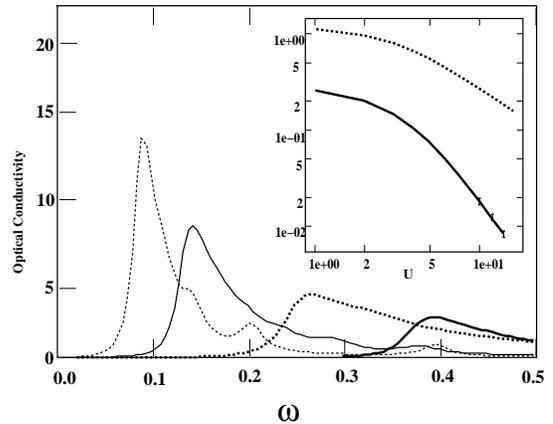}}
\caption{The optical conductivity spectra
of the periodic Anderson model for values of the interaction
$U= 0.5,1,2,3$ (right to left), keeping the hybridization
$V = 0.25$ fixed. The inset shows
the gap from the optical spectra $\Delta_c \approx \Delta_{dir}$
and the indirect gap $\Delta_{ind}$ from the local density of states
for $V=0.6$. The slopes of these curves indicate that
${V^*}^2/D  \propto \Delta_{ind}$ and $V^* \propto \Delta_{dir}$
in the strong correlation region.}
\label{fig12}
\end{figure}

The mean field theory accounts for all these observations. The low energy
behavior of the one particle Green functions of the model can be understood
as that of a non interacting system where the interaction $U$ reduces
the hybridization from its bare value $V$ to a renormalized value $V^*$
which decreases as $U$ increases. In consequence, the gap in the
optical conductivity decreases by the effect of correlations. However,
the lineshape remains approximately invariant, and merely changed
by a rescaling factor respect to the response of the non-interacting
model. This is demonstrated
by the plot of the optical conductivity
$\sigma (\omega)$ for different values of $U$ shown in
Fig.\ref{fig12}.
The optical gap $\Delta_c$ is given by the {\it direct gap} $\Delta_{dir}$
of the renormalized
band structure.
These results were obtained by IPT at $T=0$ and we checked in various cases
that the results are in excellent agreement with the ED method.

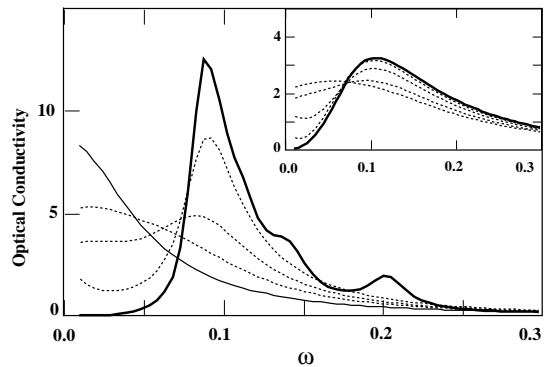
\begin{figure}
\centerline{\epsfxsize=2.9truein
\epsffile{fig24}}
\caption{The optical conductivity for the Anderson model at
$T= 0.001$ (bold), $0.005, 0.01 , 0.02$ (dotted), $0.03$ (thin).
The interaction $U=3$ and $V=0.25$.
Inset: The same quantity at $T= 0.001$ (bold),
$0.005, 0.01 , 0.02$ (dotted), $0.03$ (thin)
with lorentzian random site disorder
of width $\Gamma=0.05$.
}
\label{fig13}
\end{figure}

\begin{figure}
\centerline{\epsfxsize=2.9truein
\epsffile{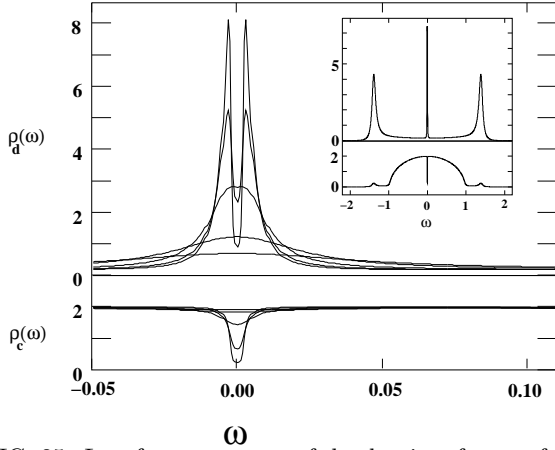}}
\caption{Low frequency part of
the density of states for the $d$ and $c-$electrons
(top and bottom) obtained from IPT at
$T$ = 0.001, 0.005, 0.01, 0.02,
0.03 for $U=3$ and $V=0.25$ (top to bottom for $d-$electrons and bottom to top
for $c-$electrons). Inset: The density of states
in the full frequency range at $T \ = \ 0.001$.
}
\label{fig14}
\end{figure}

We now consider the behavior of $\sigma(\omega)$ with temperature.
Fig.\ref{fig13} shows the optical
conductivity for different temperatures  with the parameters
$U=3$ and $V=0.25$ fixed.
The gap is essentially temperature independent. It begins to
form at $T^* \approx 0.02 \sim \frac{\Delta_c}{5}$, and is fully depleted
only at temperatures of the order of $\frac{T^*}{5}$.
We thus observe that the mean field theory is able to capture the
qualitative aspect of the experimental results that we summarized before.
This basically consists in the individualization of 3 different energy
scales: the largest corresponds to the gap of the optical spectra
$\Delta_c \sim \Delta_{dir}$, an intermediate scale $T^* \sim
\frac{\Delta_c}{5}$ where this gap starts to form and quasiparticle
features start to appear in the $DOS$ and, a third and smaller
scale $\Delta_{ind} \sim \frac{T^*}{5}$,
which corresponds to
the temperature where the optical gap gets completely depleted.
As demonstrated in Fig.\ref{fig14} where we plot the results for the
density of states, that smallest scale $\Delta_{ind}$
also indicates the temperature
below which the gap in the density of
states opens, and, thus, can be associated
to the gap measured in dc-transport experiments $\Delta_t$.

In order to
make a meaningful comparison with the experimental data, we have added the
effects of disorder by putting a lorentzian distributed random site
energy on the conduction electron band with width $\Gamma=0.05$.
The results are displayed in the inset
of Fig.\ref{fig13} and they show that the introduction of disorder
makes the overall shape of the spectra in closer agreement
with the experimental results \cite{bucher,schlesinger}
(for a discussion of the scattering involved, see
subsection \ref{henrik}).
Also, we observe that increasing
the disorder {\it reduces} the temperature $T^*$.

\begin{figure}
\centerline{\epsfxsize=2.9truein
\epsffile{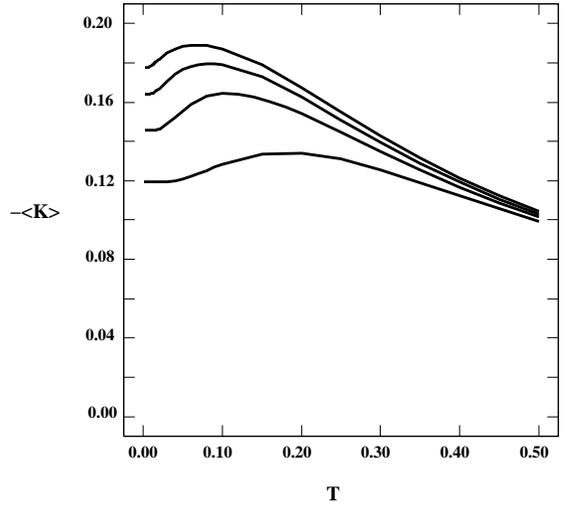}}
\caption{Expectation value of minus the kinetic energy $\langle K \rangle$
as a function of the temperature
for $U=0,2,3,4$ and $V=0.4$ (bottom to top) as
obtained from (IPT).
This quantity is directly related to the optical conductivity
sum rule. It predicts a notable {\it decrease} in the total optical
spectral weight as the temperature is {\it decreased} in the range below
the maxima.
}
\label{fig15}
\end{figure}

\begin{figure}
\centerline{\epsfxsize=2.9truein
\epsffile{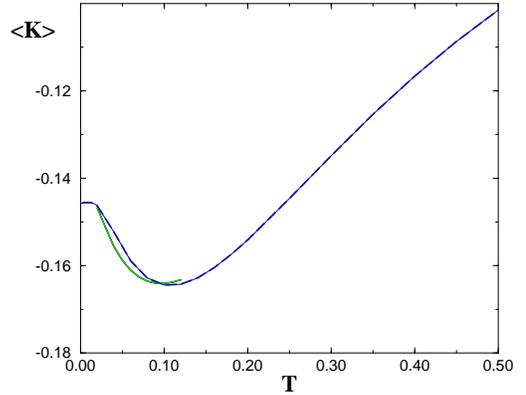}}
\caption{Comparison of the
expectation value of the kinetic energy $\langle K \rangle$ in the PAM
as a function of the temperature for $U=2$ and $V=0.4$
as obtained from IPT and ED method (dashed line and full line).
}
\label{fig15a}
\end{figure}

In the following, we  briefly
address the question of the integrated total spectral weight.
It has been noticed that the experimental results in both,
$Ce_3Bi_4Pt_3$ and $FeSi$, seem to violate the sum rule for the
spectral weight \cite{bucher,schlesinger}. However, this point has been
recently questioned, at least for the $FeSi$ compound \cite{ott}.
In order to contribute to the proper interpretation of the experimental
data, it is important to compute the kinetic energy of our model at finite
temperature, which
is directly related to the sum rule of Eq.\ref{sum}.
The results from IPT are presented in Fig.\ref{fig15} which shows the notable
dependence of the kinetic energy with temperature and interaction strength
(we plot the negative of $\langle K \rangle$ which is the quantity that
enters Eq.\ref{sum}).
In Fig.\ref{fig15a}, we plot similar results obtained with the ED algorithm
which demonstrates that the behavior captured by the IPT calculation is
indeed a true feature of the model.

As we have previously discussed for the
the Hubbard model case, the strong correlation effects that
renders $\langle K \rangle$ a function of the temperature implies
that {\it if the PAM is the relevant model for the systems at low energies},
then the results predict the behavior of the integrated optical weight
within the low frequence range.
Actually, experimental data which is inferred from the
Kramers-Kronig transformation of reflectivity measurements,
can only be reliably obtained
within a limited low frequency range of the order of a few $eV$.
The behavior of $\langle K(T) \rangle$ in Fig.\ref{fig15a} is non-monotonic.
As we increase the temperature from zero, we observe that initially
the kinetic energy decreases. This is a consequence of the electron
delocalization since the system becomes a metal
as the small gap in the density of states is filled. The kinetic energy
then goes through a minimum and starts to increase as the temperature
is further increased. This is simply due to the thermal excitation
of electrons within the single conduction band. Correlations now play
an irrelevant role as the temperature is higher than the coherence temperature
$T^*$.
When we study the behavior for different values of the interaction $U$
in Fig.\ref{fig15},
we observe that the position of the minima (maxima in this figure as
$- \langle K(T) \rangle$ is plotted), becomes smaller as
$U$ is increased. This can be understood simply as a consequence of
the renormalization of the hybridization amplitude $V \rightarrow V^*$.

In regard to the experimental situation in the Kondo insulators,
which indicate the apparent violation of the optical sum rule,
our results give a plausible qualitative
explanation for the observed
behavior. In fact, for experimental data obtained at temperatures
smaller than the size of the gap $\Delta_c$ and restricted to a finite
low frequency range (which is in fact the actual situation),
the model predicts the apparent ``disappearance''
of spectral weight as the temperature is decreased.

We should also point out that although this simple model
accounts, rather successfully, for the various energy scales,
it fails to provide an
accurate reproduction of the detailed experimental lineshape. A complete
explanation of the experimental results may need the consideration
of additional sources of scattering, as will be discussed in the
next subsection.

To finish our discussion on the gap formation in the
periodic Anderson model we shall
present the results for the size of the various gaps that are
obtained from the correlation functions.

The first study of the periodic Anderson model in large dimensions
was carried out by
Jarrell {\sl et al.} using quantum Monte Carlo \cite{mjam}.
Our spectral functions and density of states are in
general agreement with the early work in the region where the QMC and
exact diagonalization method can be compared.
A noticeable qualitative difference is that we find the spin gap to be
slightly but strictly smaller than the indirect gap when $U\ne0$.

\begin{figure}
\centerline{\epsfxsize=2.9truein
\epsffile{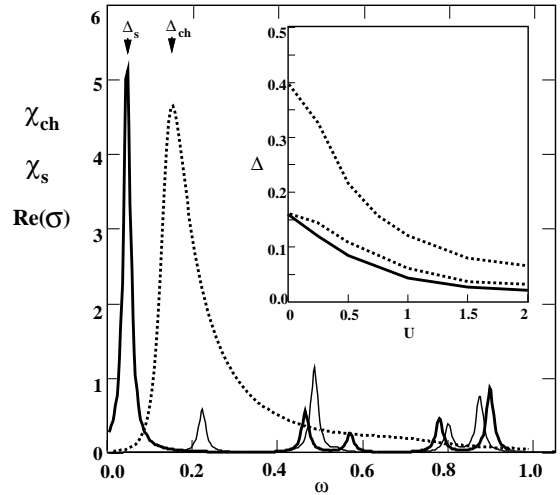}}
\caption{The local spin-spin (bold) and charge-charge (thin)
susceptibility from 7 sites ED.
The optical conductivity from IPT
(dotted). The parameters are $U = 1$ and $V=0.2$. The y-axis is in
arbitrary units.
Inset: The direct gap form IPT (upper dotted line),
the indirect gap (lower
dotted line) and the spin gap (solid line) from 8 sites ED.
The hybridization is $V=0.2$.}
\label{fig16}
\end{figure}

In Fig.\ref{fig16} we show the local spin and charge correlation functions
along with the optical conductivity which shows the qualitative agreement
with the experimental data of Ref.\onlinecite{bucher}.
We also compare in the inset the direct optical gap $\Delta_{dir}$,
the indirect gap $\Delta_{ind}$ relevant for transport properties,
and the spin gap $\Delta_s$ obtained from the spin-spin correlation function.
We find that $\Delta_{dir}$ is consistently larger than $\Delta_s$, and that
$\Delta_s$ is somewhat smaller than $\Delta_{ind}$. As expected when
$U=0$, $\Delta_s = \Delta_{ind}$, but as $U$ increases
$\frac{\Delta_s}{\Delta_{ind}}$ becomes smaller than unity and approaches
the value $1/2$ at $U \approx 2$.

\subsubsection{The scattering rate.}
\label{henrik}
In the previous section we have stressed the
qualitative success of the mean field theory of the periodic Anderson
model in connection  with the gap formation
in Kondo insulators like $FeSi$ and
$Ce_3Bi_4Pt_3$.
 In this section, however, we will show that this
approach {\em cannot} account for the large scattering rate measured in
these materials, if one does not include the effects of disorder in the model.
This is very surprising and is an indication of the limitations of
the one band periodic Anderson model for modeling these systems.

 The optical
conductivity of
Kondo insulators
  is -- except for the gap which forms at low temperatures --
almost constant over a large frequency range extending to several
 times the width of the gap.
The corresponding value of  $\sigma(\omega)$ is quite similar for all
of the materials (typically 3000 - 4000 $(\Omega cm)^{-1}$\,) and
depends only weakly on temperature.
The related scattering rate can be estimated (at $T > T^*$)
by simple Drude model
arguments: at zero frequency, we have $\sigma=\frac{ne^2\tau}{m}$.
Here, $n=a^{-3}$, where $a$ denotes the lattice constant.
 $m$ can be obtained from the kinetic energy
$\frac{p^2}{2m}\approx D$, where $p\approx\frac{2\pi\,\hbar}{a}$.
Assuming $a\approx 10^{-10}m$, the equations yield
$\hbar/\tau\sim \sigma_0^{-1} 10^3 (\Omega cm)^{-1} D$. Thus the
measured values for $\sigma_0$ imply a scattering rate which is of
the order of the bandwidth
($\frac{1}{\tau} \sim D$, assuming $\hbar=1$).
This should be compared with the scattering rate found in normal
metals like copper, which is three orders of magnitude smaller
($\frac{1}{\tau} \sim 10^{-3}D$\/).

\begin{figure}
\centerline{\epsfxsize=2.9truein
\epsffile{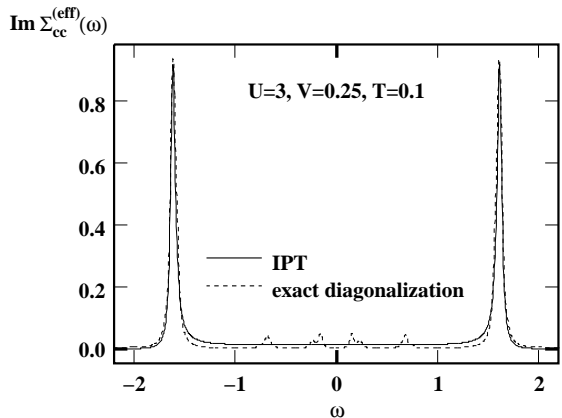}}
\caption{$\mbox{Im} \Sigma^{(eff)}_{cc}(\omega)$ for $U=3$,
$V=0.25$, and $T=0.1$ ($D=1$). The results are from
ED (dashed line) and IPT (full line).}
\label{fig29}
\end{figure}

Since  all  experiments on Kondo insulators
(and also on many Kondo metals)
observe (above the gap)
the same order of magnitude for  $\sigma(\omega)$
 one should expect that there is a common
mechanism involved.
It is reasonable to assume that the
scattering of conduction electrons
by the localized electrons in the periodic Anderson model
provides an explanation.
To address this question, we calculated the effective scattering rate.
This quantity is determined by the
effective c-electron self-energy $\Sigma_{cc}^{(eff)}$
\begin{equation}
\Sigma_{cc}^{(eff)} (\omega) =\frac{V^2}{\omega+\mu-\Sigma_{dd}(\omega)},
\end{equation}
where $\Sigma_{dd}$ is the self-energy of the localized $d-$electrons,
which enters the formula for the optical conductivity (\ref{sigma})
via $A_{\epsilon_k}(\omega)=
-2\, \mbox{Im}
\frac{1}{\omega+\mu-\epsilon_k-\Sigma^{(eff)}_{cc}(\omega)}$.
The imaginary part of $\Sigma_{cc}^{(eff)}(\omega)$ measures the scattering
involved. In Fig.\ref{fig29} we plotted this quantity for the
particle hole symmetric case, $V=0.25D$, $U=3D$,
and $T=0.1D$. Since in this section we are
not interested in the gap formation the temperature was chosen to be
well above the point where the gap starts to open ($T^*\approx0.025
D$). For comparison, the calculation was done with both ED and IPT.
It is clear from the plot that the scattering rate is much smaller
than the bandwidth $D$ and gives rise to an optical conductivity which is
smaller than the experimentally observed value by two orders of magnitude.
This result remains valid away from particle
hole symmetry and for different choices of $V$ and $U$.
If one ignores the self consistency condition, one would expect,
based on the theory of the Kondo impurity model that as the temperature
is lowered the scattering rate should grow towards the unitary limit,
$ D$ , this growth, which is expected at low frequencies and low temperatures
is preempted, in the lattice by the formation of the hybridization gap.

For $\frac{V}{U}\ll 1$ and half filling, the periodic Anderson model can be
transformed into a
Kondo lattice model by a Schrieffer-Wolff Transformation:
\begin{equation}
  H_{KL}= \frac{D}{2} \sum_{<ij>,\sigma} c^{\dagger}_{i\sigma}\,c_{j\sigma} +
             J \, \sum_{i,\sigma \sigma^{\prime}}
	     \vec{S}_i \, c^{\dagger}_{i\sigma}
            \vec{\tau}_{\sigma,\sigma^{\prime}} c_{i \sigma^{\prime}}
\end{equation}
where $J\,=\,8\,\frac{V^2}{U}$. Here, $\vec{S}_i$ describes a spin at
site $i$. For a cross-check, we also examined this hamiltonian
using the
exact diagonalization method. For $J=\frac{1}{6}\,D$, which
corresponds
to $V=0.25\, D$ and $U=3\,D$, we find at $T=0.1\,D$ a scattering rate
$\frac{1}{\tau}\sim -\mbox{Im}\Sigma_{cc}(\omega=0)\approx 7\times 10^{-3}
D$. This is again much less than required to explain the experimental
data.

We believe that the failure observed here within the present approach
is a general shortcoming
of the periodic Anderson model.
A more realistic description has to take several
crystal filed split  bands and this could increase the finite
frequency optical absortion.

\section{\bf Conclusions.}
\label{conc}
In this paper
we have illustrated how the LISA,
that becomes exact in the limit of large dimensions, can be
used to  study the physics of systems where the
local interactions are strong and play a major role. In particular
we have  demonstrated that the Hubbard and the periodic Anderson model
treated within this dynamical mean field theory can account for
the main features of the temperature dependent
transfer of spectral weight  in the  optical
conductivity spectra.
In the case of $V_2O_3$ we found that the theory is able to
account semi-quantitatively for the conductivity results in
both the metallic and insulating states. In the
former case it can also account for the topology and energy
scales of the experimental phase diagram as well as for
the unusually big values
observed in the slope of the specific heat $\gamma$.
In the latter case, the theory
seems to provide further insights in the role of the magnetic
frustration. In this regard,
we have studied in detail the predictions of the model for photoemission
spectra in the insulating phase with long range order and noted that
the present mean field theory indeed captures many aspects of
the behavior encountered in the numerical studies of the model
in low dimensions.

For the Kondo insulators,
we have seen most of the qualitative features of
the observed behavior of optical spectra with temperature
being captured in detail by our model treated in mean field
theory. In particular we identified the
 different energy scales that describe
the thermal filling of the optical gap
and how they relate to the changes in the single particle spectra.
However, we saw that the periodic Anderson model is not able to
explain the high scattering rate measured
in Kondo insulators.

We presented results for the temperature dependence
of the optical sum rule in the strongly correlated models. While in
the Hubbard case, the results
capture the qualitative change of the total spectral weight with
temperature observed in the $V_2O_3$ system, our quantitative
results on the PAM
may be relevant for the resolution
of the ``missing'' spectral weight controversy in optical experiments on
the insulators $Ce_3Bi_4Pt_3$ and $FeSi$ \cite{bucher,schlesinger,ott}.
{}From a broader perspective it has turned out to be very illuminating to
realize how the emergence of a small ``Kondo'' energy scale which
is a correlation effect results in an unusual
temperature dependence
 of the projected optical
sum rule ({\it cf.} Eq.\ref{sum})
in the Hubbard and the periodic Anderson model.
We have shown that in the former case the optical weight
increases when the temperature is reduced and the system becomes more
metallic, while in the latter it decreases, as a consequence
of the system opening a gap and becoming insulating.

We finally stress that our mean field
approach can be easily adapted  to  incorporate more realistic band
structure density of states and more  complicated unit cells.
These extensions  would allow for a more
precise quantitative description of these interesting  systems.

\acknowledgments
We acknowledge valuable discussions with V. Dobrosavljevic, A. Georges,
L. Laloux, E. Miranda,
G. Moeller, Q. Si, G. Thomas. This work was supported by the NSF
under grant DMR 92-24000.

\appendix
\section{Addition of two continued fractions.}

In this appendix we present an algorithm that allows to sum
two continued fractions into a single one. This is
necessary for the
implementation of the ED method in models with magnetic
frustration or disorder, where various Green functions
have to be averaged and the result has to be expressed as
a new continued fraction.
The details of the ED method can be found in Ref.\onlinecite{srkr}.

In the ED method an effective cluster hamiltonian $H^{n_s}$ of
$n_s$ sites is diagonalized.
At $T=0$ only the groundstate $| gs \rangle$ and the groundstate
energy $E_0$ need to be obtained, and this can be
efficiently done by the modified Lanczos method.\cite{gagliano}
The local Green function $G(\omega)$ is then obtained as a continued fraction.
Actually one needs to compute two continued fractions $G^<(\omega)$ and
$G^>(\omega)$, for
$\omega < 0$ and for $\omega > 0$ respectively.

\begin{eqnarray}
G(\omega) & = & G^{>}(\omega) + G^{<}(\omega)  \nonumber  \\
& = & \langle  gs|c \frac{1}{\omega - (H^{n_s}-E_0) +
i\delta}c^\dagger| gs \rangle \nonumber \\
& &+  \langle  gs|c^\dagger \frac{1}{\omega + (H^{n_s}-E_0) +
i\delta}c| gs \rangle
\end{eqnarray}

with

\begin{eqnarray}
G^>(\omega)=
\frac
{\langle gs|c c^\dagger| gs\rangle }
{\omega -a_0^> -\frac{b_1^{>2}}{\omega  -a_1^{>}
-\frac{b_2^{>2}}{\omega  -a_2^>-...}}} \nonumber
\\
\nonumber
\\
G^<(\omega)=
\frac
{\langle gs|c^\dagger c| gs\rangle }
{\omega -a_0^< -\frac{b_1^{<2}}{\omega  -a_1^{<}
-\frac{b_2^{<2}}{\omega  -a_2^<-...}}}
\end{eqnarray}

where $c$ and $c^\dagger$ are the operators associated with the local site
of $H^{n_s}$.
The parameters $a_i^{>/<}$ and $b_i^{>/<}$ define the continued fractions
and are obtained from the following iterative procedure,

\begin{equation}
a_i^\alpha= \langle f_i^\alpha|H^{n_s}|f_i^\alpha \rangle ,\ \ \ \ \ \
{b_i^\alpha}^2 = \frac{\langle f_i^\alpha|f_i^\alpha \rangle}
{\langle f_{i-1}^\alpha|f_{i-1}^\alpha \rangle }
\label{lan1}
\end{equation}
where $\alpha = >,<$ and
$|f_0^>\rangle = c^{\dagger} |gs \rangle $, $|f_0^< \rangle =
c |gs \rangle$ and
\begin{equation}
|f_{i+1}^\alpha \rangle = H^{n_s} |f_i^\alpha \rangle -
a_i^\alpha |f_i^\alpha
\rangle - {b_i^\alpha}^2 |f_{i-1}^\alpha\rangle
\label{lan2}
\end{equation}
and in the beginning we set $b_0^\alpha=0$.

Thus, we observe that the basis defined by the vectors
$|f_i^\alpha \rangle$ gives a {\it tri-diagonal} representation
of $H^{n_s}$ which contains the $a_i$'s along the main diagonal
and the $\sqrt{b_i}$'s along the diagonals next to the main one.
In the following we drop the index $\alpha$ to simplify
the notation. We will explicitly restore it in the final result.

Let's now address the problem of our current interest.
We assume that we have computed two Green functions
$G_\mu(\omega)$ where the index $\mu$ may label, for instance, a spin.
Our task is to obtain a new continued
fraction representation of the {\it average} Green function
$\bar{G}(\omega)= \frac{1}{2}(G_\uparrow(\omega)+G_\downarrow(\omega))$.
The more general case of a weighted average can be trivially
generalized from the present case which we consider for
simplicity.

We first note that, from the Lanczos procedure, $H^{n_s}$
has (different) {\it tri-diagonal} representations in the two sub-basis
defined by $|f_{i\mu} \rangle$ (we have dropped the $>, <$ label to
simplify notation).
The representation is basically
a matrix that contains the parameters $a's$ along the main diagonal
and the $b's$ along the two sub-diagonals.

The algorithm is as follows:
one first diagonalizes the two tri-diagonal representations of
$H^{n_s}$ by computing all the eigenvalues and eigenvectors. This
is not numerically costly since the matrices are in tri-diagonal form and
it may be done by standard methods.

An important result that can be easily demonstrated is that
the eigenvalues $\epsilon_\mu^\nu$
of the tri-diagonal matrices are the poles of their corresponding
Green functions $G_\mu(\omega)$. Furthermore,
one can also show that

\begin{equation}
G_\mu(\omega)=\sum_{\nu=1}^M
{\frac{({v^{\nu}_\mu})^2}{\omega-\epsilon^\nu_\mu}}
\end{equation}
where $v^{\nu}_\mu$ are the first component of the $M$
eigenvectors of the tri-diagonal matrices.

Thus, from the definition of the Green function, one
immediately recognizes that the vector
\{$v^1_\mu,v^2_\mu,...,v^M_\mu$\} is nothing but
$c^\dagger_\mu| gs \rangle$ expressed in a basis where $H^{n_s}$ is diagonal
(which is a sub-basis of the given sector's Hilbert space).

The final step consists in writing the hamiltonian in the basis
direct product of the two sub-basis, which, of course, will
also be a diagonal representation of $H^{n_s}$; and then bring it to
its tri-diagonal representation through the steps
(\ref{lan1}), (\ref{lan2}) starting from the vector defined by (restoring
the $ >, <$ label)

\begin{eqnarray}
|f_0^> \rangle & = &
(c^{\dagger}_\uparrow + c^{\dagger}_\downarrow ) |gs \rangle
\nonumber  \\
& = &  \vec{v}_\uparrow \oplus \vec{v}_\downarrow  \nonumber  \\
& = &  \{v^1_\uparrow,v^2_\uparrow,...,v^M_\uparrow,
v^{M+1}_\downarrow,v^{M+2}_\downarrow,...,v^{2M}_\downarrow\}
\end{eqnarray}

Thus, the newly determined $a_i^>$'s and $b_i^>$'s that result
from this last step are the parameters
of the continued fraction representation of $\bar{G}^>(\omega)$
(the parameters for $\bar{G}^<$ are obtained in a completely analogous
manner).

\end{document}